\documentclass[a4paper,11pt]{article}
\pdfoutput=1 

\usepackage{jcappub} 

\usepackage[T1]{fontenc} 

\newcommand{\dd}{\mathrm{d}}
\newcommand{\sgn}{\mathrm{sgn}}

\title{\boldmath Chiral Magnetic Effect in Protoneutron Stars and Magnetic Field Spectral 
Evolution}

\author{G\"unter Sigl}
\author{and Natacha Leite}
\affiliation{II. Institut f\"ur Theoretische Physik, University of Hamburg\\
Luruper Chaussee, 149, 22761 Hamburg, Germany}

\emailAdd{guenter.sigl@desy.de}
\emailAdd{natacha.leite@desy.de}

\abstract{We investigate the evolution of the chiral magnetic instability in a 
protoneutron star and compute the resulting magnetic power and helicity spectra. The instability may act during the early cooling phase
of the hot protoneutron star after supernova core collapse, where it can contribute to the buildup of magnetic fields of strength up to the order of $10^{14}$ G. The maximal field strengths generated by this instability, however, depend considerably on the temperature of the
protoneutron star, on density fluctuations and turbulence spectrum of the medium. At the end of the hot cooling phase the magnetic field tends to be concentrated around the submillimeter to cm scale, where it is subject to slow resistive damping.}

\begin{document}
\maketitle
\flushbottom

\section{Introduction}
\label{sec:intro}
The origin of the magnetic field strengths observed in neutron stars 
and magnetars (highly magnetized neutron stars \cite{magnetar}) up to 
$10^{15}$ G  \cite{Kouveliotou, catalog} is still under debate. The most 
popular explanations include adiabatically compressed fossil fields of the parent star and dynamo generated fields \cite{1504.08074}.
More recently it has been suggested that the magnetic field of magnetars is related to a chiral asymmetry of particles,
produced during the core collapse of supernovae \cite{1402.4760}. An imbalance in the number of right- and 
left-handed fermions was previously studied in the context of QCD plasmas 
\cite{warringa, 1402.4174} as well as applied to the early universe as a 
possibility to explain the generation  and evolution  of cosmological magnetic 
fields \cite{prl108, vachaspati}. This so-called chiral magnetic effect or chiral magnetic instability
was also suggested to account for the observed kicks that accelerate neutron stars \cite{1410.3833}.

In a pure electron-positron plasma, the chiral magnetic instability does not allow 
the growth of seed magnetic fields \cite{1405.3059}, but in the presence of neutrinos
an electroweak plasma with neutrino-antineutrino asymmetries was found to be able to amplify
magnetic fields to interesting scales for neutron stars 
\cite{1311.5267, 1409.1463}. Another crucial ingredient to take into account are the spin flip interactions
due to the finite electron mass which violates chirality. This tends to decrease the 
asymmetry between left- and right-handed electrons faster than it is created by electroweak processes
\cite{1409.3602}. In addition, it was claimed that the chiral asymmetry in the 
forward scattering amplitude of electrons off nuclei due to the electroweak interaction can create a magnetic field instability in the same 
way that the chiral asymmetry does, but which acts on much longer time scales 
and is not washed out by chirality-flipping processes \cite{1410.6676,1503.04162}.

It is not yet clear if the chiral magnetic instability can transfer sufficient energy stored in chiral fermions into magnetic field energy
to give a significant contribution to the magnetic fields inferred for neutron stars and magnetars.
In the present paper we aim to model the chiral magnetic effect in this environment,
review the underlying assumptions proposed for this mechanism to work and understand its physical implications. We solve the evolution equations for the chiral chemical potential, the chemical potential of the background species on which electrons scatter, the magnetic energy and the magnetic helicity power spectra. This allows us to estimate the conditions for which magnetic fields can be amplified in a neutron star. We show that a seed magnetic field can be amplified to small scales shortly after the collapse in the core of hot stars through the chiral magnetic effect and that for its surface or cooled down neutron stars this mechanism is not effective to generate the strong magnetic fields observed.

The remainder of the paper is structured as follows: In Sect.~\ref{sec:framework} we summarize the conditions in a protoneutron
star relevant for magnetic field evolution and set up the basic modified MHD equations. In Sect.~\ref{sec:solutions} we solve the
evolution equations, estimate the maximal magnetic field strength and discuss assumptions and uncertainties. We summarize our
results and conclude in Sect.~\ref{sec:conclusions}. Throughout the paper we will use Gaussian natural units, $c_0=\hbar=k_B=1$, and the electric
permittivity and magnetic permeability of the vacuum are set to $\epsilon_0 = 1/(4\pi)$ and $\mu_0 = 4\pi$, respectively.

\section{Framework and basic equations}
\label{sec:framework}

\subsection{Thermodynamics of a protoneutron star}
\label{sec:thermodynamics}
Immediately after core collapse, a protoneutron star reaches 
temperatures of the order of tens of MeV in its core. To a given temperature
corresponds a chemical potential $\Delta \mu = \mu_n - \mu_p = \mu_e-\mu_\nu$ \cite{burrows86}, 
that can be used together with the fact that neutrinos are trapped inside the 
core at this stage, such that the lepton fraction $Y_L$ is temporarily conserved.
The relation $Y_L n_B = n_e + n_\nu$, where $n_B = n_n + n_p$ is the baryon number density, and electric neutrality, $n_e = n_p$,
allows us to estimate the number densities and chemical potentials of the particle species involved.

When a massive star collapses protons are converted into neutrons by capturing left-handed electrons $e_L+p \rightarrow n + \nu_{e_L}$, producing an asymmetry between the number of left- and right-handed electrons $N_5 \equiv N_L - N_R$. Such electroweak reactions are known as URCA processes and their emissivity is \cite{lattimer}
\begin{equation}
\epsilon_{\rm URCA} = \frac{457 \pi}{10080}(1+3g_A^2)\cos^2\theta_C G_F^2 m_n m_p \mu_e T^6\,,
\end{equation}
where $g_A \simeq 1.26$ is the axial-vector coupling of the nucleon,  $\theta_C\approx 0.24$ is the Cabbibo angle, $G_F = 1.166\times 10^{-5}$ GeV$^{-2}$, $m_n$ and $m_p$ are the masses of the neutron and proton, respectively and $\mu_e$ is the electron chemical potential.
Since $\mu_e\gg T$, the rate of electron capture is then
\begin{equation} \label{eq:Gw}
\Gamma_w = \frac{\epsilon_{\rm URCA}}{\mu_e Y_L n_B}\,.
\end{equation}
If the URCA processes are not in thermodynamic equilibrium with the inverse reactions, an asymmetry $N_5$ can build up. This is the case, for example, if neutrinos escape the neutron star, which occurs when their mean free path is larger than the neutron star radius. This condition is met beyond the neutrino sphere or when enough time has passed for the star to cool down to the point when it becomes transparent to neutrinos, roughly 10 seconds after 
collapse. 

For the typical momenta of the particles in the core of a protoneutron star electron capture takes place, while in the crust or when proton and electron concentrations are low momentum conservation highly suppresses electron capture and an additional particle is required to absorb momentum, such as another proton or neutron. If $p_{F,n}> p_{F,e}+p_{F,p}$, the previous rate is modified to \cite{modURCA}
\begin{equation} \label{eq:Gwm}
\Gamma_w^{\rm mod} \simeq \frac{11513 \pi}{120960}\alpha_\pi^2G_F^2\cos\theta_C g_A^2 \frac{T^8}{Y_L n_B}\,,
\end{equation}
where $\alpha_\pi \approx 15$ is the pion-nucleon fine structure constant. 

During the hot initial phase, the charge carriers in the neutron star are semi-degenerate. There are no simple equations for the conductivity
for this case. In the degenerate limit the conductivity is given by the following expression~\cite{cond},
\begin{equation} \label{eq:cond}
 \sigma \simeq 1.5\times 10^{45}\left(\frac{\rm K}{T}\right)^2\left(\frac{\rho_p}{10^{13} {\; \rm g\, cm}^{-3}}\right)^{3/2} \; {\rm s}^{-1}\,,
\end{equation}
with $\rho_p$ the proton density. In the relativistic non-degenerate high temperature limit the conductivity would be dominated by the
pair plasma and be of the order of the temperature~\cite{hotcond} which is lower than the obtained through \eqref{eq:cond}. However, it turns out that the final magnetic field is insensitive to the conductivity in this range of values because the magnetic field grows on timescales much shorter than the dynamical timescale of the system and then saturates. Therefore, for our calculations we will use eq.~(\ref{eq:cond}) for the conductivity, which is closer to the relevant conditions in the protoneutron star, and we will also assume the temperature $T$ and the conductivity $\sigma$ to be constant.

Even though in the core of a collapsing supernova electrons are relativistic, the fact 
that they are massive suggests that we should not take them as strictly chiral 
particles, since the amplitude of a positive helicity component for a 
left-chiral state is approximately $(E+m_e-p)/(E+m_e) \simeq (m_e/E)$.
This means that there is a probability $(m_e/E)^2$ that a scattering electron of 
a certain chirality flips into the opposite chirality state -- either by 
Rutherford scattering, electron-electron scattering or Compton scattering -- 
which tends to decrease $N_5$. Rutherford scattering dominates in this case, which 
allows us to write the chirality-flipping rate as ($E \sim T$)  \cite{1409.3602}
\begin{equation} \label{eq:flip}
 \Gamma_f \simeq \frac{e^4 m_e^2}{48\pi^3 \mu_e}\left[\ln\frac{12\pi^2 
T}{e^2(3T+\mu_e)}-1\right]\,,
\end{equation}
where $e$ is the electron charge and $m_e$ is the electron mass.

\subsection{Evolution equations}

When a chiral imbalance is present, such as the one originated by electron capture, the Adler-Bell-Jackiw anomaly implies that a current 
\begin{equation}
j_5^\mu \equiv  \bar{\psi}\gamma^\mu\gamma_5\psi = j_L^\mu-j_R^\mu
\end{equation}
will be induced, whose partial derivative, instead of vanishing, is related to 
the Chern-Simons current through
\begin{equation}
\partial_\mu j_5^\mu  = \frac{g^2}{32\pi^2}F^\alpha_{\mu \nu}\tilde{F}^{\alpha, \mu \nu} 
= \partial_\mu K^\mu\,.
\end{equation}
With $N_5 = \int \dd^3\mathbf{r }\bar{\psi}\gamma_5\psi$ and $N_{CS} \equiv \int \dd ^3 \mathbf{r } K^0$ space integration implies
the conservation relation 
\begin{equation} 
\frac{\dd}{\dd t}\left(N_5 - N_{CS} \right)= 0\,.
\end{equation}
Thus the Chern-Simons number of the electromagnetic field connects the chiral 
asymmetry to the magnetic helicity
\begin{equation} \label{eq:cons}
\frac{\dd}{\dd t}\left(N_5 - \frac{e^2}{4\pi^2}\mathcal{H} \right)= 0\,, \; 
\mathcal{H} = \int \dd ^3 \mathbf{r }\, \mathbf{B}\cdot\mathbf{A}\,,
\end{equation}
with the magnetic field $\mathbf{B}$ and vector potential $\mathbf{A}$. 

Maxwell's equations are then modified by the introduction of a current contribution 
\begin{equation}
 \mathbf{j}_5=-\frac{e^2}{2\pi^2}\mu_5 \mathbf{B}  
\end{equation}
in the presence of a chiral imbalance, with chiral chemical potential $\mu_5 \equiv(\mu_L - 
\mu_R)/2$. This affects the magnetohydrodynamics (MHD) equation that now takes the form
\begin{equation} \label{eq:Bevol}
\partial_t \mathbf{B} = \mathbf{\nabla}\times(\mathbf{v}\times \mathbf{B}) + 
\eta \Delta \mathbf{B} - \frac{2e^2}{\pi}\eta \mu_5 
\mathbf{\nabla}\times\mathbf{B}\,,
\end{equation} 
with $\eta = 1/(4\pi\sigma)$ being the resistivity. In the following we will neglect the velocity field ${\bf v}$
so that our subsequent analysis applies in the plasma rest frame, provided that the velocity field is
sufficiently smooth, which we elaborate in appendix \ref{app:turb}. It will be left to future work to investigate under which conditions this is a good
approximation in the presence of turbulence and other contributions to the velocity such as rotation.

In Fourier space and introducing the expansion of the magnetic field into a 
left- and right-handed part 
\begin{equation}\label{eq:hel_b}
\mathbf{\tilde{B}(k)} = b_\mathbf{k}^+\mathbf{h}_\mathbf{k}^+ + 
b_\mathbf{k}^-\mathbf{h}_\mathbf{k}^- \,, \hspace{0.5cm} 
\mathbf{h}_\mathbf{k}^\pm \equiv  \frac{1}{\sqrt{2}}\left(\mathbf{e}\pm 
i\frac{\mathbf{k}}{k}\times \mathbf{e}\right)\,,
\end{equation} 
where $\mathbf{e}$ is an arbitrary unit vector perpendicular to $\mathbf{k}$, 
one can then rewrite \eqref{eq:Bevol} as
\begin{equation} \label{eq:bevol}
\partial_t  b_\mathbf{k}^\pm = - \eta k \left(k \pm 
\frac{2e^2}{\pi}\mu_5\right)b_\mathbf{k}^\pm\,.
\end{equation}

The magnetic field energy density and helicity density can be written in terms 
of the power spectra in Fourier space, $M_k\equiv 
k^3|\mathbf{\tilde{B}(k)}|^2/2$ and $H_k \equiv -4\pi i k 
[\mathbf{k}\times\mathbf{\tilde{B}(k)}]\cdot \mathbf{B^*(k)}$, as
\begin{equation}
\begin{split}
\rho_m &=\frac{1}{V} \int \dd^3 \mathbf{r}\frac{\mathbf{B}^2(\mathbf{r})}{8\pi}=\frac{M_k}{V}\int_0^\infty \dd
\ln k M_k \,, \\
h &=\frac{1}{V} \int_0^\infty \dd \ln k H_k \,.
\end{split}
\end{equation}
Combining \eqref{eq:hel_b} and \eqref{eq:bevol} and multiplying with the magnetic field complex 
conjugate, the power spectra evolution is given by
\begin{equation}\label{eq:Mevo}
\partial_t \rho_m = -\frac{\eta}{V} \int \dd \ln k k^2 \left(2M_k + 
\frac{e^2}{2\pi^2}\mu_5 H_k \right)\,, 
\end{equation}
\begin{equation} \label{eq:Hevo}
 \partial_t h = -\frac{\eta}{V} \int \dd \ln k (2k^2 H_k + 32e^2\mu_5 
M_k)\,,
\end{equation}
where $V$ is the volume. We can translate particle number into chemical potential using
\begin{equation} \label{eq:n5tomu5}
N_5 = \frac{V}{3\pi^2}  \mu_5 \left(\mu_5^2 + 3\mu_e^2 + \pi^2 T^2 \right)\,,
\end{equation} 
which can be approximated to linear order in $\mu_5$ to $N_5 = c(T, 
\mu_e)V\mu_5$, with
\begin{equation}
c(T, \mu_e) = \frac{\mu_e^2}{\pi^2}+ \frac{T^2}{3}\,. 
\end{equation}
To express the evolution of the chiral chemical potential we take into account 
the processes described by \eqref{eq:flip} and \eqref{eq:cons} that affect the 
number of left- and right-handed particles, resulting in
\begin{equation} \label{eq:mu5evo}
\partial_t \mu_5 = \frac{e^2}{4\pi^2 c(T, \mu_e)}\partial_t h -2\Gamma_f 
\left(\mu_5 - \mu_{5,b}\right)\,,
\end{equation}
where $\mu_{5,b}$ is the equilibrium value of $\mu_5$ in the absence of 
resistivity. This term represents an effective chemical potential generated by 
the interactions of electrons with background species such as neutrinos that act as sources of the 
asymmetry, thus containing the term proportional to \eqref{eq:Gw}.
We can roughly estimate it by considering the number $N_b$ of background 
particles, such that the processes that change $N_5$, neglecting the magnetic field 
contribution for now, can be written as
\begin{equation}
\partial_t N_5 = \pm\Gamma_w N_b- 2\Gamma_f N_5\,.
\end{equation}
Comparing this with \eqref{eq:mu5evo}, results in
\begin{equation} \label{eq:Nb}
N_b = 2Vc(T, \mu_e) \frac{\Gamma_f}{\Gamma_w}|\mu_{5,b}|\,.
\end{equation} 
Furthermore, we approximate the background particles at the temperatures of the core of a protoneutron star to be non-degenerate relativistic fermions with $g_b$ degrees of freedom. This is plausible since the chiral asymmetry mostly results from URCA processes involving chiral neutrinos whose chemical potential is at most of the order of the temperature which would give rise to order one corrections.
From thermodynamics we can then relate the number of background particles with the temperature using $N_b = 3\zeta(3)Vg_bT^3/(2\pi)^2$, yielding
\begin{equation}
 \label{eq:mu5b}
|\mu_{5,b}|=\frac{3\zeta(3)}{8\pi^2}g_b \frac{\Gamma_w}{\Gamma_f}\frac{T^3}{c(T, \mu_e)}\,.
\end{equation}
This effective background chemical potential can also include a possible contribution from
the difference of the forward scattering amplitudes of left- and right-handed 
electrons on nucleons, that was considered in the form of an effective potential $V_5$ in
\cite{1410.6676} and \cite{1503.04162}. Requiring energy conservation of the combined system
consisting of the magnetic field, the chiral asymmetry in the electro sector and the background particles
implies an evolution equation for $\mu_{5,b}$ of the form of 
\begin{equation} \label{eq:muLevo}
\partial_t \mu_{5,b} = \Gamma_{ w}\frac{\mu_5}{\mu_{5,b}}(\mu_5 - \mu_{5,b})\,,
\end{equation}
as we will see below.

\subsection{Magnetic field amplification} \label{sec:Bfield}

The initial magnetic field of a protoneutron star, which for example can result from adiabatic compression of
the stellar seed field during collapse, is affected by the 
chiral magnetic effect in a way which depends on the scales we are interested 
in. From \eqref{eq:bevol} for growing modes we obtain the condition
\begin{equation} \label{eq:k5}
k< \frac{2e^2}{\pi}|\mu_5| \equiv k_5(\mu_5)\,,
\end{equation}
such that magnetic field modes with $k>k_5$ decay due to resisitivity.
For $k<k_5$ the magnetic field mode with the same sign for helicity and 
$\mu_5$ also decays while the mode with helicity signs opposite to $\mu_5$ grow. 
All magnetic field modes are damped due to 
finite resistivity with the resistive damping rate $\Gamma_r = \eta k^2$ whereas
the growth/decay rate due to the chiral instability is given by
\begin{equation} \label{eq:Gchi}
 \Gamma_\chi(k) = \frac{2e^2}{\pi} \eta  k |\mu_5| = \frac{k_5}{k}\Gamma_r \,.
\end{equation}
The maximal total growth rate 
$\Gamma_{\rm tot}= \Gamma_\chi - \Gamma_r$ occurs at $\Gamma_{\rm max} =\eta 
k_5^2/4$, which corresponds to the wavenumber $k_5/2$.\\
Let us analytically analyze the expected behavior of $\mu_5$ by setting
$\partial_t \mu_5 =0$ in \eqref{eq:mu5evo}. If we normalize $H_k$ to the maximal 
value that the helicity can take, $H_{\rm max}(k)= 8\pi M_k/k$, this gives
\begin{equation}
\tilde{\mu}_5 = \frac{\Gamma_f \mu_{5, b} - 2\eta e^2[\pi V c(T, 
\mu_e)]^{-1}\int \dd \ln k k M_k (H_k/H_{\rm max})}{\Gamma_f+4\eta e^4[\pi^2c(T, 
\mu_e)]^{-1}\rho_m}\,.
\end{equation}
When the magnetic field is negligible, $\tilde{\mu}_5 \simeq \mu_{5, b}$, and 
modes smaller than $k_5$ grow exponentially at the rate $\Gamma_{\rm tot}$. 
The magnetic field terms begin to dominate when $\rho_m \gtrsim \pi^2c(T, 
\mu_e)/(4\eta e^4)\Gamma_f$. In this limit the flipping rate is negligible 
compared to the magnetic field induced rate and the instability yields
\begin{equation}
\tilde{\mu}_5 \simeq -\frac{\pi}{2e^2 \rho_m} \int \dd \ln k k 
\frac{M_k}{V}\frac{H_k}{H_{\rm max}}\,.
\end{equation}
In general $\tilde{\mu}_5 \neq \mu_{5, b}$ and the  terms of \eqref{eq:mu5evo} compensate each other. 
As a consequence the magnetic helicity density will change linearly with time at a rate
\begin{equation}
\partial_t h\simeq \frac{8\pi^2 c(T, \mu_e)}{e^2}\Gamma_f(\mu_5-\mu_{5, b})\,.
\end{equation}
The fact that we have almost maximal helicity implies that also the magnetic 
energy density changes linearly with time, either growing or decreasing 
according to the sign of $(\mu_5-\mu_{5, b})/h$. The helicity would be constant 
only if $\Gamma_f=0$ or if $\mu_5=\mu_{5, b}$. Assuming that the magnetic 
energy is concentrated around a characteristic scale $k_0 = k_5(\tilde{\mu}_5)$ 
and that the helicity is maximal and has opposite sign to $\tilde{\mu}_5$ implies that
also $\rho_m$ will be constant. In this case the chiral magnetic 
instability reaches saturation, where the growth and damping rates compensate each other.

\subsection{Energy balance} \label{sec:energy}

By definition of a chemical potential the energy $E_5$ associated with the chiral asymmetry
is given by $ \dd E_5 = \mu_5 \dd N_5$ and with $E_5=0$ for $\mu_5=0$ results in
\begin{equation} \label{eq:E5}
\rho_5 = \frac{E_5}{V} = \frac{c(T, \mu_e)}{2} \mu_5^2\,.
\end{equation}
Differentiating this with respect to time, using \eqref{eq:mu5evo} with the 
helicity normalized to its maximal value and inserting $k_5$, we obtain
\begin{equation} \label{eq:rho5evo}
\partial_t \rho_5 = -2\eta \int \dd \ln k \frac{M_k}{V} \left(k_5 k \sgn(\mu_5) 
\frac{H_k}{H_{\rm max}} + k_5^2\right)- 2c(T, \mu_e)\Gamma_f\mu_5(\mu_5-\mu_{5, 
b})\,.
\end{equation}
We can also estimate the change in magnetic energy that a finite $\mu_5$ can induce by 
using that the instability produces maximally helical fields and that, as we have seen, 
the growth has its peak at $k_5/2$. Thus $\dd E_m \simeq 
k_5|\dd\mathcal{H}|/(8\pi)$ and using \eqref{eq:cons} gives $\dd E_m 
\simeq V c(T, \mu_e) \mu_5 \dd \mu_5$. If there is an initial chiral asymmetry 
$\mu_{5i}$, the increase in magnetic energy density is given by
\begin{equation} 
\Delta \rho_m \simeq \frac{ c(T, \mu_e)}{2}(\mu_{5i}^2- \mu_5^2)\,.
\end{equation} 
The total energy density
\begin{equation}
\rho_{\rm tot}= \rho_5+ \Delta\rho_m \simeq \frac{c(T, \mu_e)}{2}\mu_{5i}^2\,,
\end{equation}
then only depends on the initial value $\mu_{5i}$ which implies that the 
maximal increase in magnetic energy density obeys $\Delta\rho_m \leq \rho_{\rm tot}$.

Eq. \eqref{eq:Mevo} can be rearranged into
\begin{equation}
\partial_t \rho_m = -2\eta \int \dd \ln k k^2 \frac{M_k}{V}\left(1 + 
\frac{k_5}{k}\sgn(\mu_5) \frac{H_k}{H_{\rm max}} \right)\,,
\end{equation} 
which together with \eqref{eq:rho5evo} provides the rate of change of the total energy
\begin{equation} \label{eq:Etotevo}
\begin{split}
\partial_t \rho_{\rm tot} &= \partial_t \rho_m + \partial_t \rho_5 \\ 
&= -2\eta \int \dd \ln k \frac{M_k}{V} \left[(k -k_5)^2 +2k_5 k\left(1+ 
\sgn(\mu_5) \frac{H_k}{H_{\rm max}} \right)\right] \\
& - 2c(T, \mu_e)\Gamma_f\mu_5(\mu_5-\mu_{5, b})\,.
\end{split}
\end{equation}
The term proportional to $\mu_{5,b}$ is responsible for the energy exchange with 
external particles and we see that apart from it, since the integrand of 
\eqref{eq:Etotevo} is non-negative, the energy decreases due to chirality flips 
and finite resistivity. It is strictly conserved only for the case $\tilde{\mu}_5 = 
\mu_{5, b}$, and if the magnetic energy is concentrated in the mode $k_5$ and 
helicity is maximal with opposite sign to $\mu_{5, b}$, which are the same conditions 
mentioned above for $\partial_t \rho_m =0$.

\begin{figure}[tbp]
\centering 
\includegraphics[width=.47\textwidth]{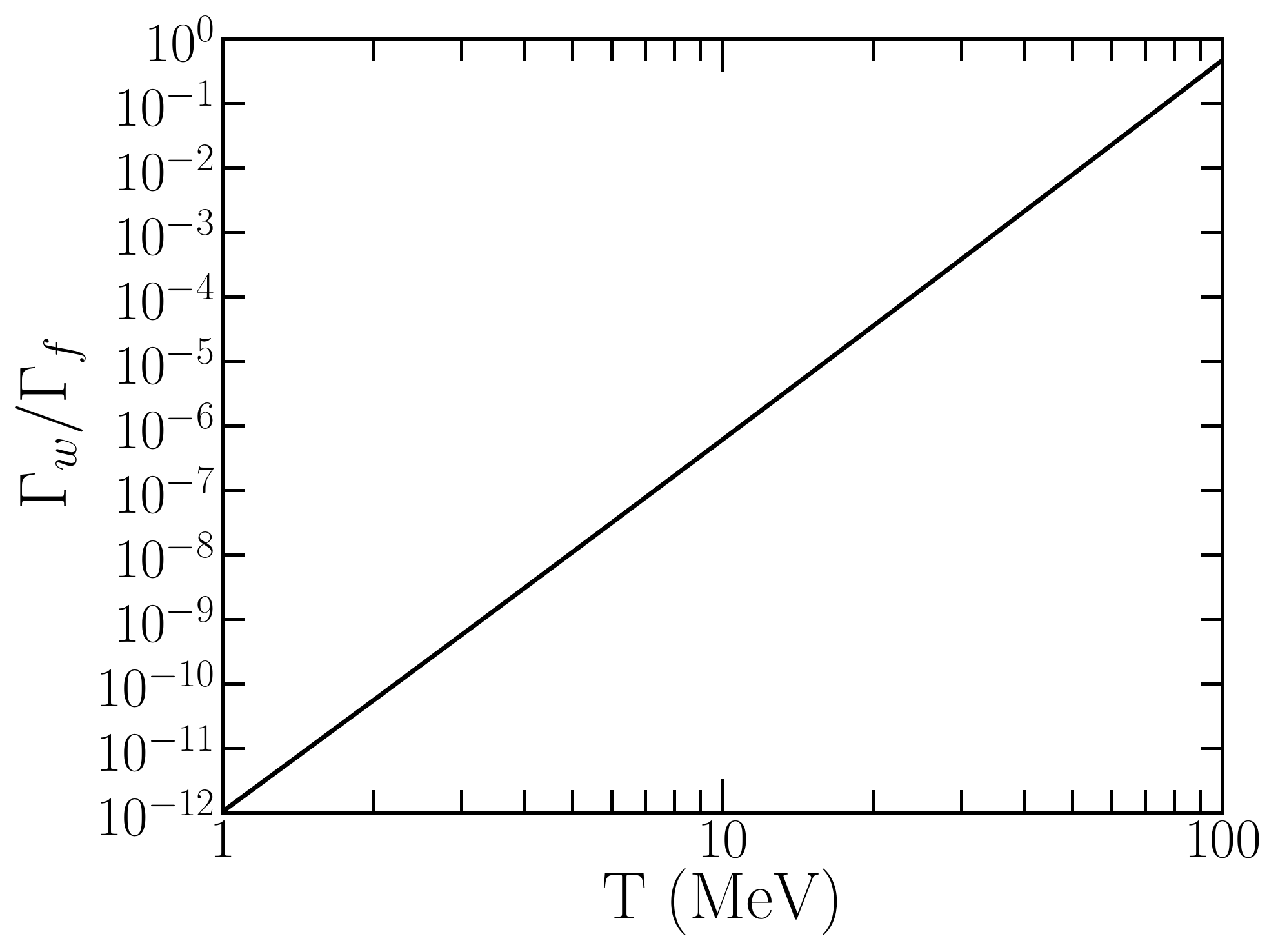}
\hfill
\includegraphics[width=.47\textwidth]{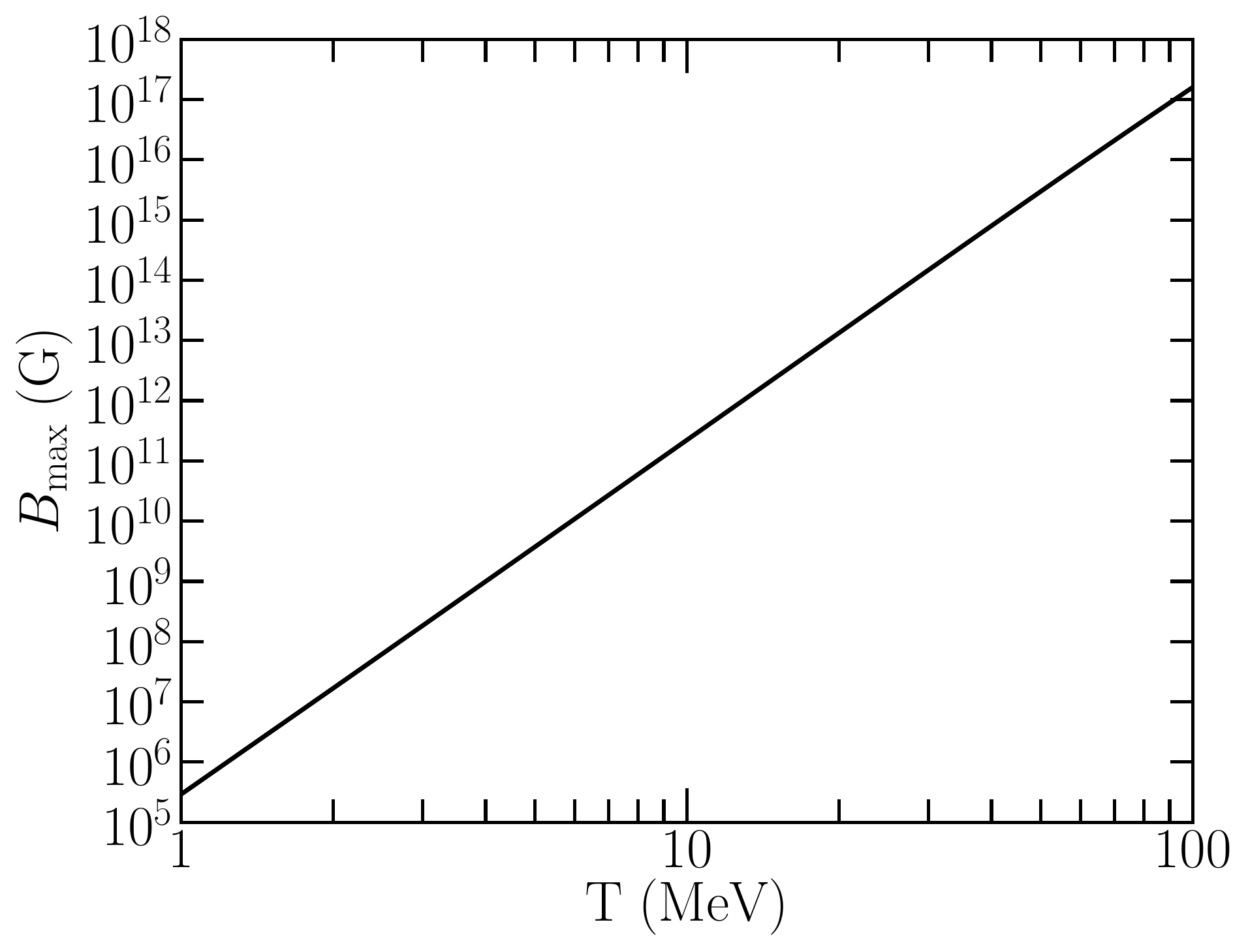}
\caption{\label{pic:ratioBmax}  Left panel: The ratio $\Gamma_w/\Gamma_f$ obtained using \eqref{eq:Gw} and \eqref{eq:flip}.
Right panel: Estimate maximum magnetic field amplification $\displaystyle B_{\rm max} = \sqrt{8\pi\rho_m^{\rm max}}$ due to the
chiral magnetic instability as a function of temperature. Logarithmic units to base 10.}
\end{figure}

Similarly to the energy associated with the chiral particles, the energy density $\rho_b$ associated with the background 
species is given by $\dd E_b=\mu_{5, b}\dd N_b$. Using \eqref{eq:Nb}, this gives
\begin{equation} \label{eq:rhob}
\rho_b = \frac{E_b}{V}= c(T, \mu_e)\frac{\Gamma_f}{\Gamma_w}\mu_{5, b}^2\,.
\end{equation}
The initial chemical potential of the background species, $\mu_{5, bi}$, gives 
us a measure of the maximal energy that can be transferred into magnetic energy 
density
\begin{equation} \label{eq:maxrhom}
\Delta \rho_m \lesssim c(T, \mu_e)\frac{\Gamma_f}{\Gamma_w} \mu_{5, bi}^2\,.
\end{equation}
For the interactions to conserve energy one has to set $\partial_t \rho_b = -\partial_t 
\rho_5$ in the absence of magnetic fields and, computing the time derivative of \eqref{eq:rhob} and using 
\eqref{eq:rho5evo}, this yields the evolution equation for the background 
species \eqref{eq:muLevo}. This shows that $\mu_{5,b}$
changes typically with the rate $\Gamma_w$ and that will be in equilibrium for $\mu_5 = 
\mu_{5,b}$ when the magnetic field is concentrated around $k_5(\mu_{5,b})$ and 
has maximal helicity of sign opposite to $\mu_{5,b}$, given by \eqref{eq:mu5b}. 
Fig.~\ref{pic:ratioBmax} shows the dependence of the ratio $\Gamma_w/\Gamma_f$ on the temperature for a protoneutron star.

Once the energy contribution \eqref{eq:maxrhom} is added to \eqref{eq:Etotevo}, the only remaining source of
energy change is  resistive damping.
Inserting \eqref{eq:mu5b} into \eqref{eq:maxrhom}, we obtain
\begin{equation} \label{eq:rhomax}
\rho_m^{\rm max} = \left(\frac{3\zeta(3)}{8 
\pi^2}g_b\right)^2\frac{\Gamma_w}{\Gamma_f}\frac{T^6}{ c(T, \mu_e)}\simeq  8.3\times10^{-3}\,\frac{\Gamma_w}{\Gamma_f}\frac{T^6}{c(T, \mu_e)}\,, 
 \end{equation}
where in the last expression the degrees of freedom of the background species 
were taken as $g_b=2$. \eqref{eq:rhomax} can be used to 
predict the maximal field strength generated for a given temperature and is shown in fig.~\ref{pic:ratioBmax}. We can 
see that the maximal magnetic field amplification strongly increases with temperature.

Up to now we assumed that $\mu_5$ evolves only in time, neglecting the possible contribution of its spatial evolution, as studied by Ref.~\cite{Boyarsky:2015faa}. There it was shown that the system can be unstable with respect to growing
  inhomogeneous modes of $\mu_5$ if $k^2/4 > 3e^4 B^2/(8\pi^4T^2)$ \cite{Boyarsky:2015faa}. If we consider the fastest growing mode $k_5/2$ and the maximum magnetic field strength generated from \eqref{eq:maxrhom}, we can estimate whether the inhomogeneity of $\mu_5$ plays a role or not in our case. We find then that the solution of \eqref{eq:mu5evo} will be stable if $\Gamma_w/\Gamma_f < 4/\pi$, which is verified for the temperature values we are interested in, as is clear from the left-hand side of Fig.~\ref{pic:ratioBmax}. This implies that for our purposes $\mu_5(x,t)\equiv \mu_5(t)$ is a safe and justified assumption. 

\subsection{Density fluctuations}
\label{sec:fluct}
Neutrinos of energy $E_\nu \simeq 3T$ are trapped at the high temperatures in the protoneutron star core, where densities easily reach $n_B=2n_0$, with $n_0\simeq1.7\times 10^{38}$ cm$^{-3}$ the nuclear matter number density. The mean free path
for absorption by a neutron is \cite{naoki}
\begin{equation}
 \ell_{\rm abs}  \simeq 4.5 \times 10^6\left(\frac{n_0}{n_B}\right)^{2/3}\left(\frac{10{\, \rm MeV}}{T}\right)^{4}\left[\left(\frac{E_\nu}{T}\right)^4+10\pi^2\left(\frac{E_\nu}{T}\right)^2+9\pi^4\right]^{-1} \; {\rm cm}
\end{equation}
and the mean free path for scattering with a neutron is
\begin{equation}
 \ell_{\rm sca}  \simeq 10^4\left(\frac{n_0}{n_B}\right)^{1/3}\left(\frac{10{\, \rm MeV}}{E_\nu}\right)^{2}\frac{10{\, \rm MeV}}{T} \; {\rm cm}.
\end{equation}
The absorption mean free path is more important than scattering and yields $\simeq1.5\,$m for $T=20$ MeV and $\simeq10\,$cm
for $T=40 $ MeV. The scattering mean free path yields $\simeq1\,$m at 20 MeV and $\simeq14\,$cm at 40 MeV. These typical temperatures will be used in the following section.

To take into account the fact that the interior of a young neutron star is turbulent we consider the existence of density fluctuations $\delta \rho $ relative to the average density $\rho$. These density perturbations seem to amount to at least 25\% \cite{Mao:2015nxa}. If the scale of the fluctuations is smaller than the neutrino mean free path, locally these regions can amplify a seed magnetic field, since the neutrinos stream freely on that scale such that the URCA processes and their reverse processes are not in thermal equilibrium and the rate of production of chiral imbalance $\mu_5$ will be of the order of the direct URCA rates $\Gamma_w$. To study the influence of the fluctuations somewhat more quantitatively we introduce an effective creation rate of chiral imbalance $\Gamma_w^{\rm eff} = \Gamma_w \delta \rho/\rho$ which is a rough estimate of the difference of the absorption and emission rates
of left chiral electrons due to the electroweak URCA interactions.
Rewriting the chiral asymmetry equilibrium value \eqref{eq:mu5b} and the characteristic wavenumber
of the instability \eqref{eq:k5} in terms of the density fluctuations, we have
\begin{equation}\label{eq:mu5k5}
 |\mu_{5,b}| = \frac{3\zeta(3)}{8\pi^2}g_b\frac{\delta \rho}{\rho} \frac{\Gamma_w}{\Gamma_f}\frac{T^3}{c(T, \mu_e)} , \hspace{1 cm} k_5^{-1} = \frac{4\pi^3 c(T, \mu_e)}{3e^2\zeta(3)g_b}\left(\frac{\delta \rho}{\rho}\frac{\Gamma_w}{\Gamma_f}T^3\right)^{-1}\,.
\end{equation}
From \eqref{eq:rhomax}, the resulting maximal field amplification with respect to the density fluctuations then becomes
\begin{equation} \label{eq:Bmaxdelta}
 B_{\rm max}\simeq \frac{3\zeta(3)g_b}{[8\pi^3c(T, \mu_e)]^{1/2}}\left(\frac{\delta \rho}{\rho}\frac{\Gamma_w}{\Gamma_f}\right)^{1/2}T^3 \simeq 7.2\left(\frac{\delta \rho}{\rho}\frac{\Gamma_w}{\Gamma_f}\right)^{1/2}\frac{T^3}{c(T, \mu_e)^{1/2}}\,.
\end{equation}

\section{Solutions of the evolution equations}
\label{sec:solutions}
We now apply the previous treatment to the core and the neutrino sphere of a protoneutron star. In a core collapse
supernova with a progenitor mass $\sim 8M_\odot$, as described by \cite{raffelt}, shortly after core collapse the lepton fraction is $Y_L \simeq 0.3$.
For the chemical potential difference we consider the two realistic cases  $\Delta \mu =  80$ MeV and 60 MeV, which 
correspond to temperatures of $\simeq 40$ MeV and 20 MeV, respectively, for a core density of $2n_0$, as before. We can then compute the number density and
chemical potential of each species as described in Sect.~\ref{sec:thermodynamics} which results in the electron chemical potential $\mu_e \simeq 260$ MeV and the proton densities $\rho_p \simeq 1.3\times10^{14} {\rm \, g \, cm}^{-3}$ for 40 MeV and $1.2\times10^{14} {\rm \, g \, cm}^{-3}$ for 20 MeV. From this
the conductivity is obtained from eq.~\eqref{eq:cond}.

We solve the system of eq.~\eqref{eq:Mevo}, \eqref{eq:Hevo}, \eqref{eq:mu5evo} and 
\eqref{eq:muLevo} for 90 wavenumber modes spanning from $k_{\rm max}=2k_5$ to $k_{\rm 
min}= 10^{-4}k_5$, with constant width in $\log_{10}k$. The time scale used is normalized to the resistive damping time 
of the instability
\begin{equation} \label{eq:tdamp}
t_{\rm damp}=\Gamma_\chi^{-1}(k_5)=
\frac{2}{\eta k_5^2}=\frac{32\pi^6c(T, \mu_e)}{9\zeta(3)^2e^4g_b^2\eta}\left(\frac{\delta \rho}{\rho}\frac{\Gamma_w}{\Gamma_f}T^3\right)^{-2}\,.
\end{equation}
We first analyze the case for which the density fluctuations are close to the average density $\delta \rho \sim \rho$ and then consider that $\delta \rho$ is lower than $\rho$ by one order of magnitude. After the initial 10 seconds of its life, the neutron star becomes transparent to neutrinos and the magnetic field amplification can be estimated by putting $\delta\rho/\rho=1$.

\begin{figure}[tbp]
\centering 
\includegraphics[width=.47\textwidth]{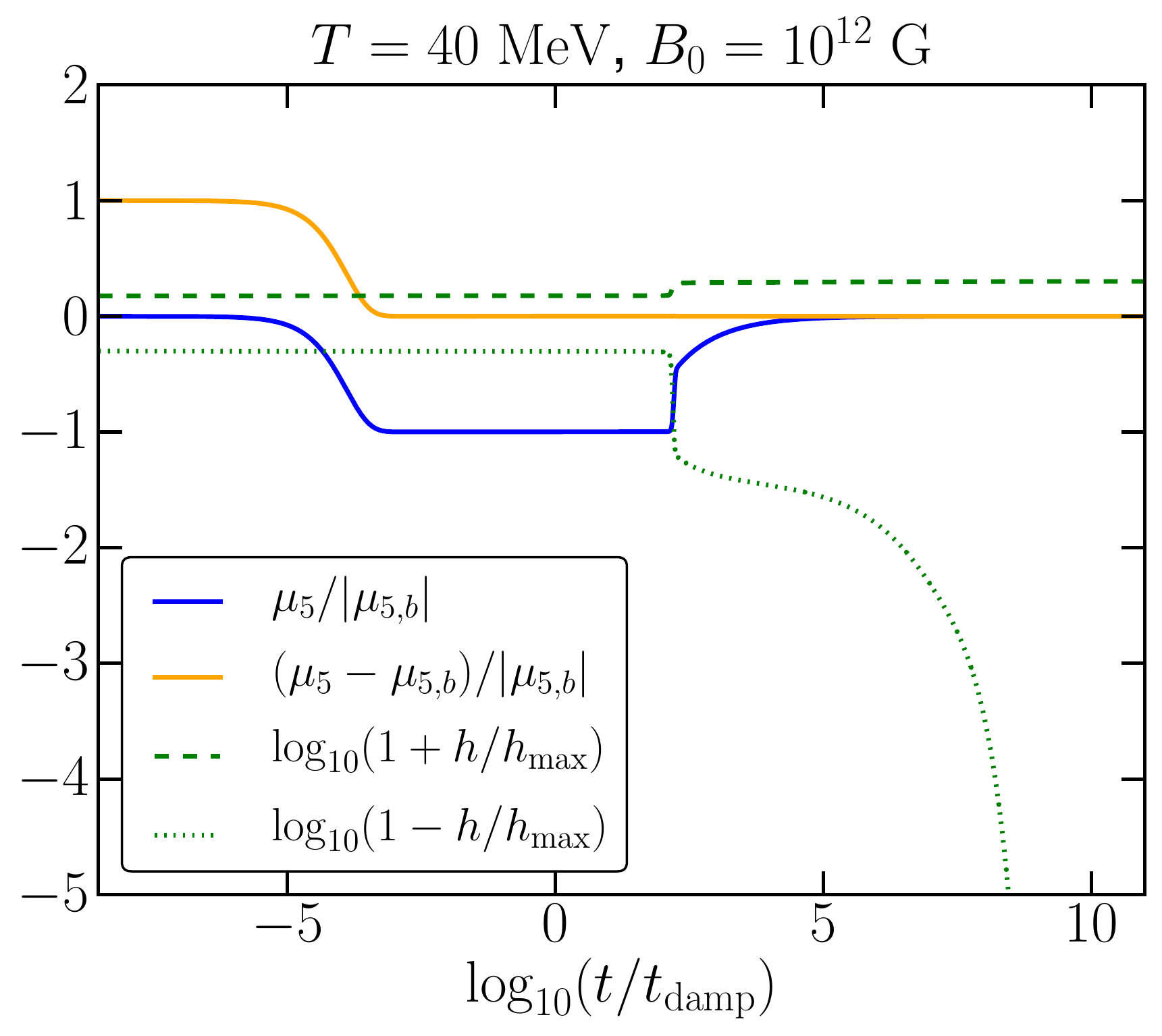}
\hfill
\includegraphics[width=.47\textwidth]{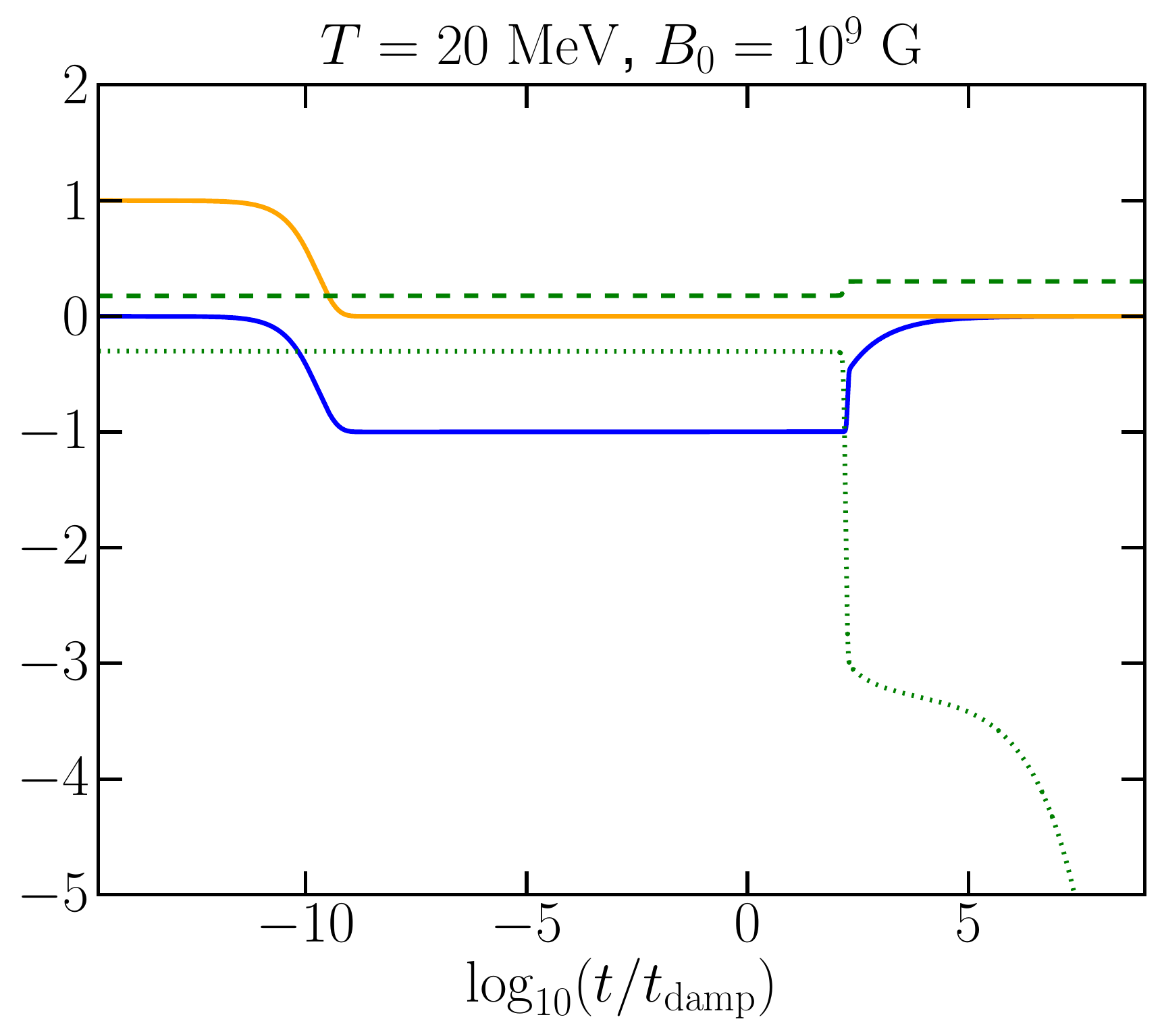}
\caption{ \label{pic:muh} Time evolution of the chiral chemical potential normalized 
to the equilibrium value, $\mu_5/|\mu_{5,b}|$, relative difference of the chiral 
chemical potential to the equilibrium value, $(\mu_5-\mu_{5, b})/|\mu_{5, b}|$ and, in logarithmic units, relative deviation of
the helicity density from its maximal and minimal value, $1\pm h/h_{\rm max}$. The left panel is for a temperature of $T=40$ MeV and seed field
$B_0=10^{12}$ G, and the right panel is for $T=20$ MeV and a seed field of $B_0=10^{9}$ G.}
\end{figure}  

\begin{figure}[tbp]
\centering 
\includegraphics[width=.47\textwidth]{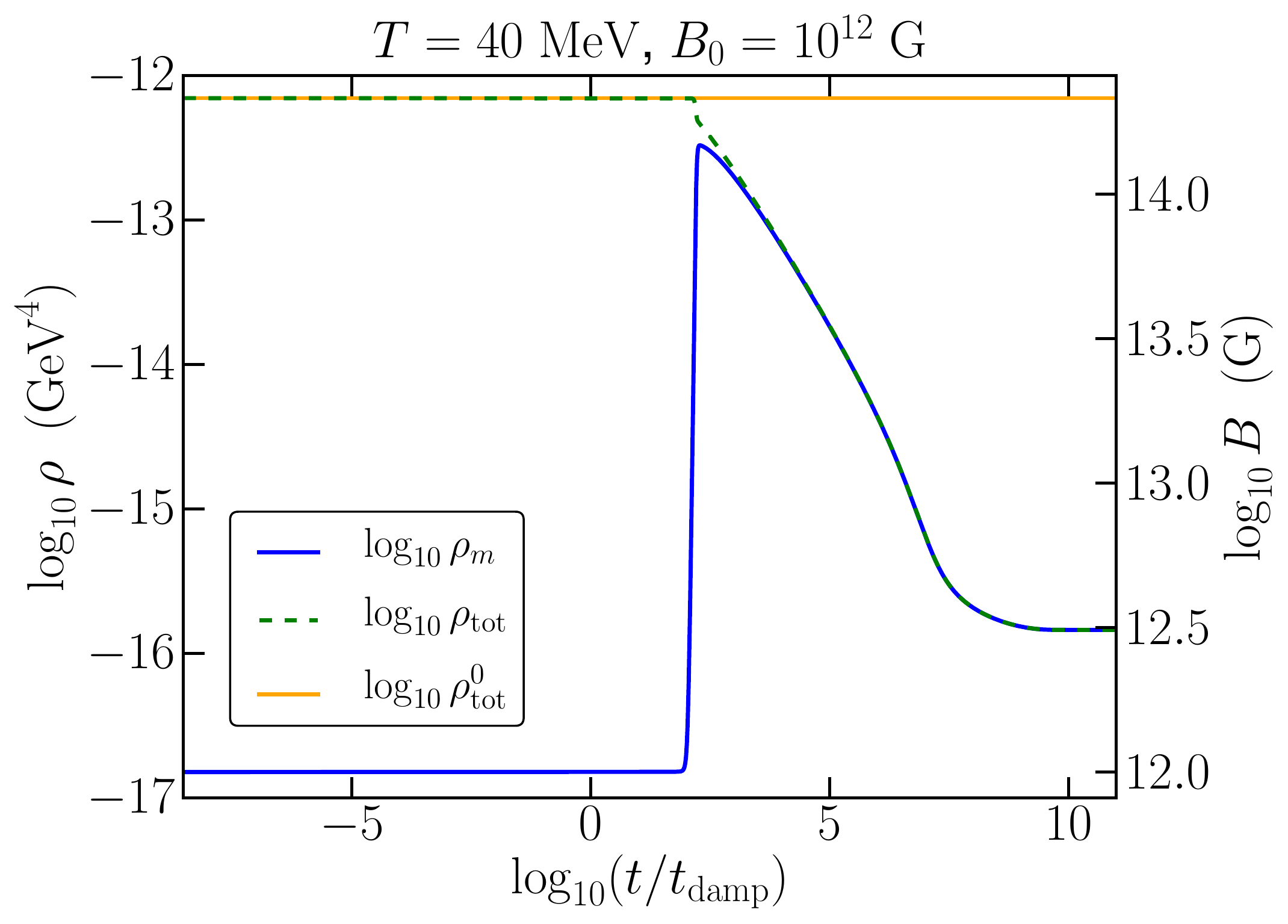}
\hfill
\includegraphics[width=.47\textwidth]{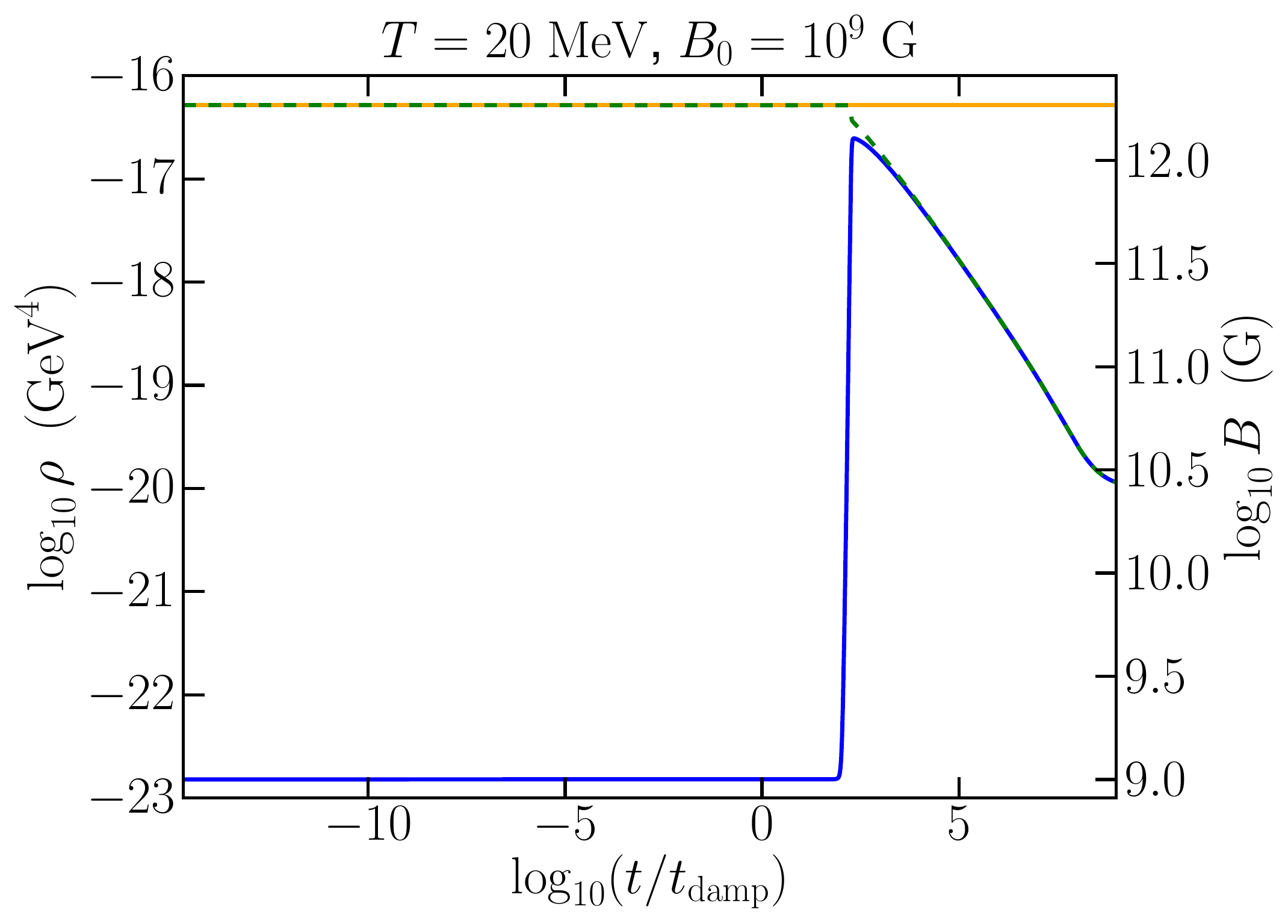}
\caption{\label{pic:rho}  Time evolution of the magnetic energy density $\rho_m$ and total energy density $\rho_{\rm tot}$. Also shown
is the initial total energy density which limits the maximal magnetic energy density that can be reached by the instability.
In the left panel $T=40$ MeV and in the right panel $T=20$ MeV.}
\end{figure}  

\begin{table}[tbp]
\centering
\begin{tabular}{| c| c | c | c | c | c | c|}
\hline
\rule{0pt}{2.5ex}  T (MeV)  & $|\mu_{5,b}|$ (MeV) & $\Gamma_w^{-1}$ (s) & $\Gamma_f^{-1}$ (s) & $t_{\rm damp}$ (s) & $k_5^{-1}$ (cm) & $B_{\rm max}$ (G) \\ \hline
\rule{0pt}{2.5ex}    40 & $2\times 10^{-3}$ & $1\times10^{-9}$ & $3\times10^{-12}$ & $6\times 10^{-8}$ & $3\times10^{-6}$ & $8\times 10^{14}$\\ \hline
\rule{0pt}{2.5ex}    20 & $4 \times 10^{-6}$ & $9\times10^{-8}$ & $3\times10^{-12}$ & $0.04$ & $1\times 10^{-3}$ & $1\times 10^{13}$ \\
   \hline 
 \end{tabular}
\caption{\label{tab:val} Equilibrium chiral asymmetry \eqref{eq:mu5k5}, chiral asymmetry creation and depletion rates \eqref{eq:Gw} and \eqref{eq:flip}, respectively, damping time \eqref{eq:tdamp} and characteristic scale \eqref{eq:mu5k5} values for $\delta \rho/\rho=1$. Maximal magnetic field amplification computed using \eqref{eq:Bmaxdelta}.}
\end{table}

In fig.~\ref{pic:muh} the evolution of the chiral magnetic instability with a vanishing initial value for the chiral chemical potential
is shown and table~\ref{tab:val} contains the respective $|\mu_{5,b}|$,  $\Gamma_w^{-1}$,  $\Gamma_f^{-1}$, $t_{\rm damp}$ and  $k_5^{-1}$ values taking  $\delta \rho/\rho=1$. 

From \eqref{eq:Bmaxdelta}, the maximum field strength that can be reached by the instability is independent of the initial magnetic seed field.
Its strong temperature dependence reflects the fact that $\Gamma_w$ has a stronger dependence on temperature than $\Gamma_f$, 
which means that the higher the temperature, the higher the magnitude of $\mu_5$ and the sooner the chiral magnetic instability develops and 
subsequently damps. The values predicted for the maximal field strength \eqref{eq:Bmaxdelta} are listed in the last column of table~\ref{tab:val}. For the examples we chose magnetic seed fields  $B_0=10^{12}$ G for $T=40$ MeV and $B_0=10^{9}$ G for $T=20$ MeV to illustrate how the magnetic field amplification is limited by energy conservation when the magnetic energy gets close to the maximum allowed value. In fig.~\ref{pic:rho}, for $T=40$ MeV the magnetic energy density grows steeply up to close to the total energy of the system,
corresponding to $B_{\rm max}\simeq 1\times 10^{14}$ G, within a few $\mu$s and subsequently the magnetic field decreases by a factor of more than 10 within a few seconds. For $T=20$ MeV, the magnetic field grows within about 4 seconds up to $B_{\rm max}\simeq 1\times 10^{12}$ G.

The total energy density 
\begin{equation}
\rho_{\rm tot}\simeq \frac{T^2}{6}\left(\mu_5^2+2\frac{\Gamma_f}{\Gamma_w}\mu_{5,b}^2\right)+\rho_m\,,
\end{equation}
includes the energy density $\rho_5$ from \eqref{eq:E5}, the energy corresponding to the background particles coupling to
the chiral electrons from \eqref{eq:rhob} which ensures that the total energy due to the scattering terms is conserved, and the magnetic 
energy density $\rho_m$. The initial total value is not exceeded and, as predicted, $\rho_{\rm tot}$ decreases only due to
resistive damping following the dissipation of $\rho_m$. The number of wavenumber modes considered is sufficiently large
for the decay of the magnetic  energy to correspond to a smooth curve. The decay changes $\rho_m$ 
roughly linearly with time, as expected from the discussion in Sect.~\ref{sec:Bfield}.

The chiral asymmetry is built up through the capture of left-handed electrons until an equilibrium with the spin-flip processes is reached at $\mu_{5} =  \mu_{5,b}$. When the magnetic field starts to be amplified, the term in 
\eqref{eq:mu5evo} proportional to the magnetic helicity will eventually dominate. At this point 
the asymmetry $\mu_5$ will start to decrease as chiral energy is transferred into magnetic energy.

The curve $(\mu_5 - \mu_{5,b})/|\mu_{5,b}|$ in fig.~\ref{pic:muh} is very 
close to zero after the equilibrium value is reached, which implies that when the 
magnetic field terms begin to dominate the evolution of $\mu_5$ and $\mu_{5,b}$ 
occurs in lockstep. The chiral chemical potential grows and reaches equilibrium at a value close 
to $\mu_{5,b}$ until the magnetic field term starts to dominate and 
depletes the chiral asymmetry. The scattering of electrons, covered by the term 
proportional to $\mu_{5,b}$ does not allow for $\mu_5$ to be replenished.

The magnetic helicity density, normalized to the maximal value $h_{\rm max}(k)=(8\pi/V) (M_k/k)$,
depends on the mode 
considered and maintains its initial value, here simply chosen as $h^0=h_{\rm max}/2$, until the amplification of the magnetic field makes it either grow to its maximum or decay, if the sign between helicity and $\mu_5$ is the opposite or equal, respectively.
Thus, fields amplified by the chiral instability turn into maximally helical fields. Furthermore, once
magnetic field growth sets in the evolution is essentially independent of the initial helicity.

\begin{figure}[tbp]
\centering
\includegraphics[width=.47\textwidth]{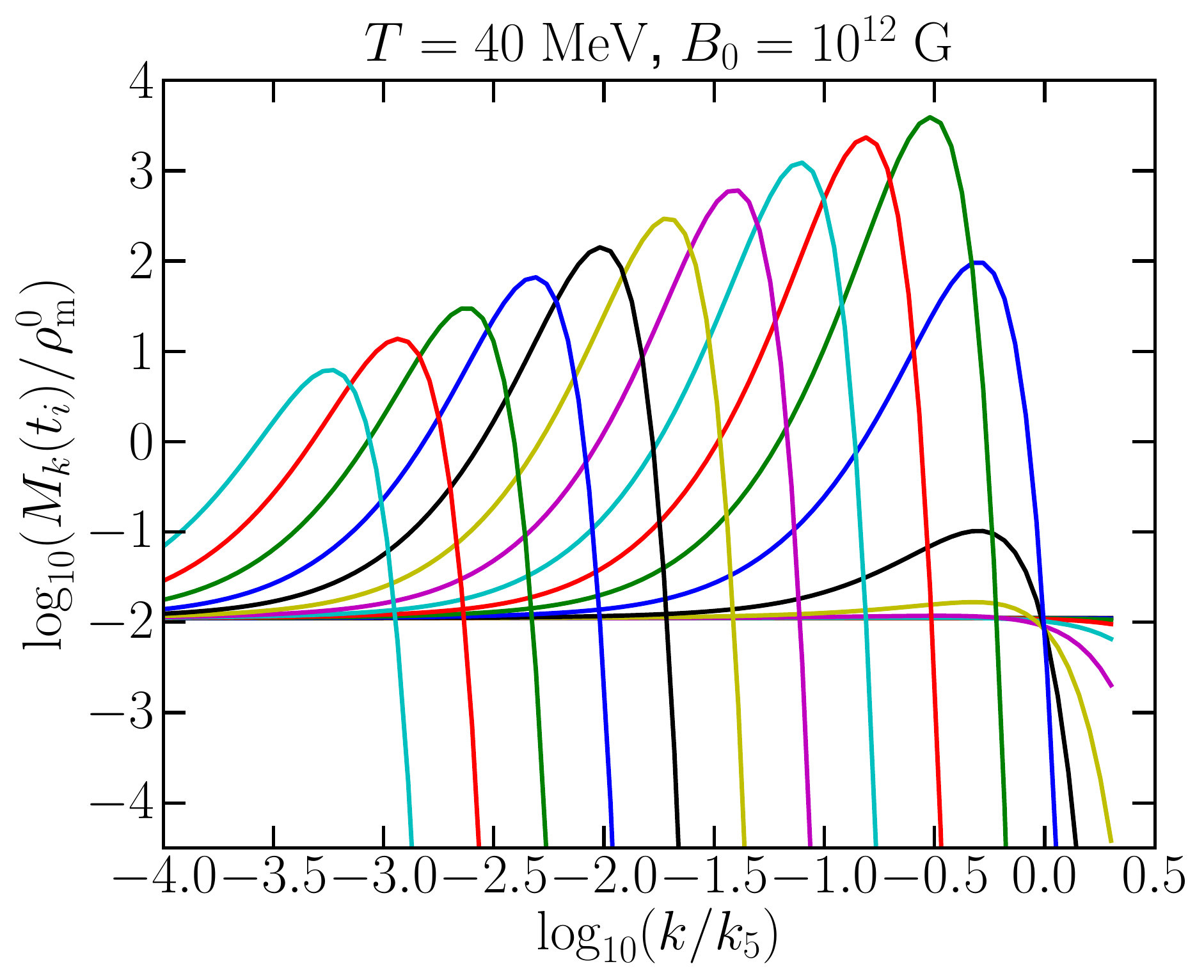}
\hfill
\includegraphics[width=.47\textwidth]{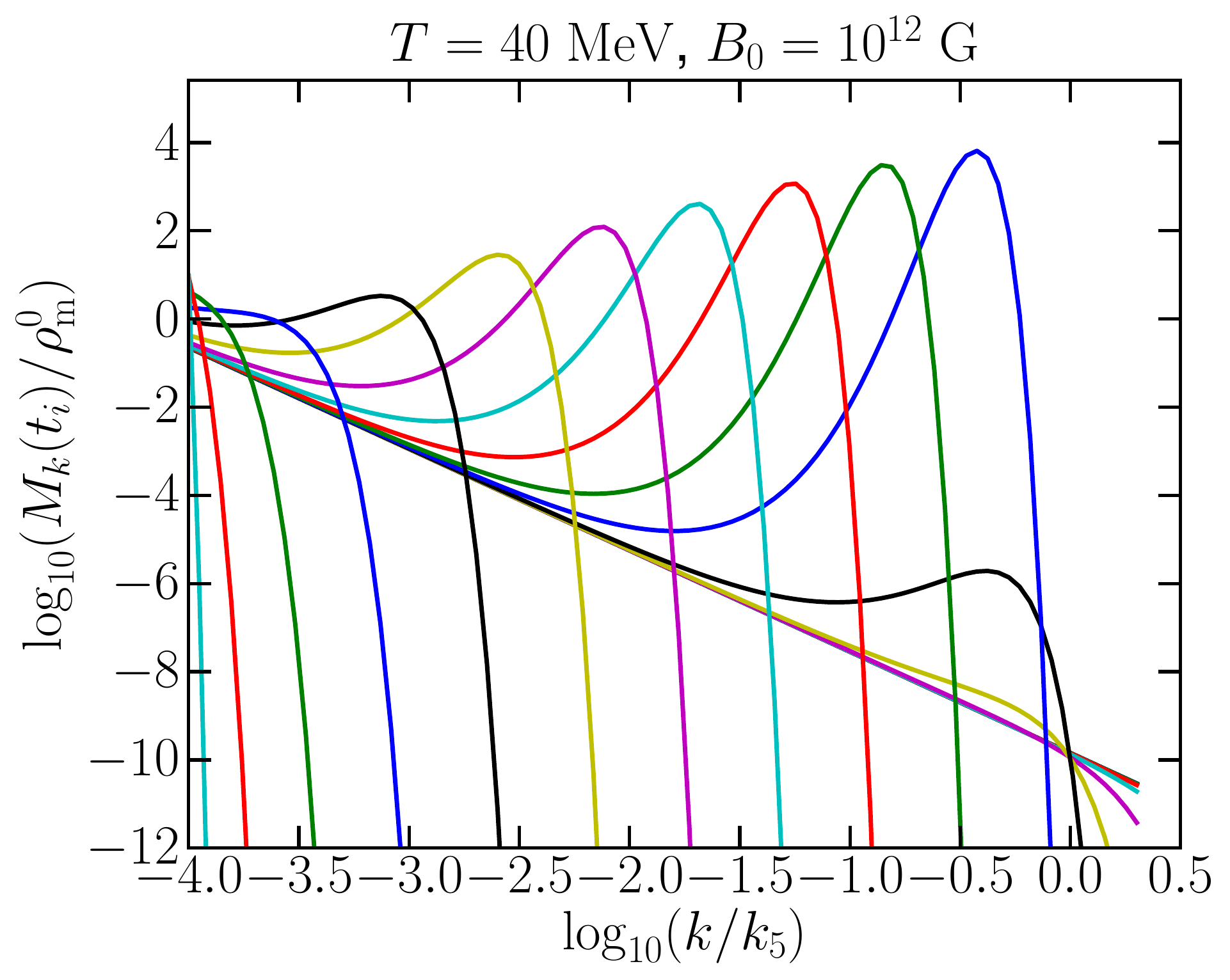}
\caption{\label{pic:spectra}  Time evolution of the magnetic field power spectrum normalized to the initial magnetic energy density,
$M_k/\rho_m^0$, as a function of wavenumber $k$ normalized to $k_5$. The power spectra are shown for equally 
spaced intervals in the logarithm of time between $t=t_{\rm damp}$ and $t=10^8t_{\rm damp}$, for $T=40$ MeV. Left panel:
Initially flat power spectrum. Right panel: Initial power spectrum has a Kolmogorov distribution.}
\end{figure}

It is also interesting to compute how the magnetic field power spectrum evolves with 
time. Fig.~\ref{pic:spectra} shows the time evolution of the magnetic field power spectrum
for a flat and a Kolmogorov initial spectrum. As expected, the final magnetic field power spectrum is not very sensitive to
the initial magnetic field power spectrum. The magnetic field power spectrum peaks at wavenumbers close to $k_5/2$,
while it decays with time due to resistive damping for $k>k_5$.
Since $k_5$ is proportional to the evolving chiral chemical potential $\mu_5$, see \eqref{eq:k5}, which
decreases for $\log_{10}(t/t_{\rm damp})>2$, first steeply and then smoothly,
see fig.~\ref{pic:muh}, with growing time the peak in the magnetic power spectrum moves to smaller $k$.
This is reflected in fig.~\ref{pic:spectra} which also shows that the total magnetic energy grows exponentially 
for times $\simeq10^2\,t_{\rm damp}\lesssim t\lesssim10^3\,t_{\rm damp}$, then saturates and 
gets damped for subsequent times.

In the simulations of the magnetic field power spectra we considered the most relevant modes $k$ for magnetic amplification: The peak of the magnetic field power spectrum yields a maximal growth for $k_5/2$ and taking into account much smaller wavenumber modes up to the size of the neutron star radius, $k\sim 1/(10 \, \text{ km})$, does not significantly change our results.

We can also estimate the time dependence of $k_5$ in the damping regime:
Eq.~\eqref{eq:Mevo} shows that amplification stops and resistive damping sets in when $2\eta k_5^2t\sim1$.
Therefore, we expect the scaling
\begin{equation}
 k_5 \sim k_5^0\left(\frac{t_0}{t}\right)^{1/2}\,, \;  \mu_5 \sim \mu_5^0\left(\frac{t_0}{t}\right)^{1/2}\,,
\end{equation}
where $k_5^0$ and $\mu_{5}^0$ are the values of $k_5$ and $\mu_5$, respectively, at $t\simeq t_{\rm damp}$.
Fig.~\ref{pic:spectra} allows us to estimate the length scale at which the power spectrum peaks after the magnetic field growth ends. For $T=40$ MeV we obtain $k^{-1}\simeq 0.1$ mm while for $T=20$ MeV we find $k^{-1}\simeq 3$ cm.
This is significantly smaller than the neutrino mean free path discussed in Sect.~\ref{sec:fluct}. Therefore, in the presence of
significant density fluctuations on these length scales we expect that magnetic field growth due to the chiral magnetic instability
is possible.
\begin{figure}[tbp]
\centering 
\includegraphics[width=.47\textwidth]{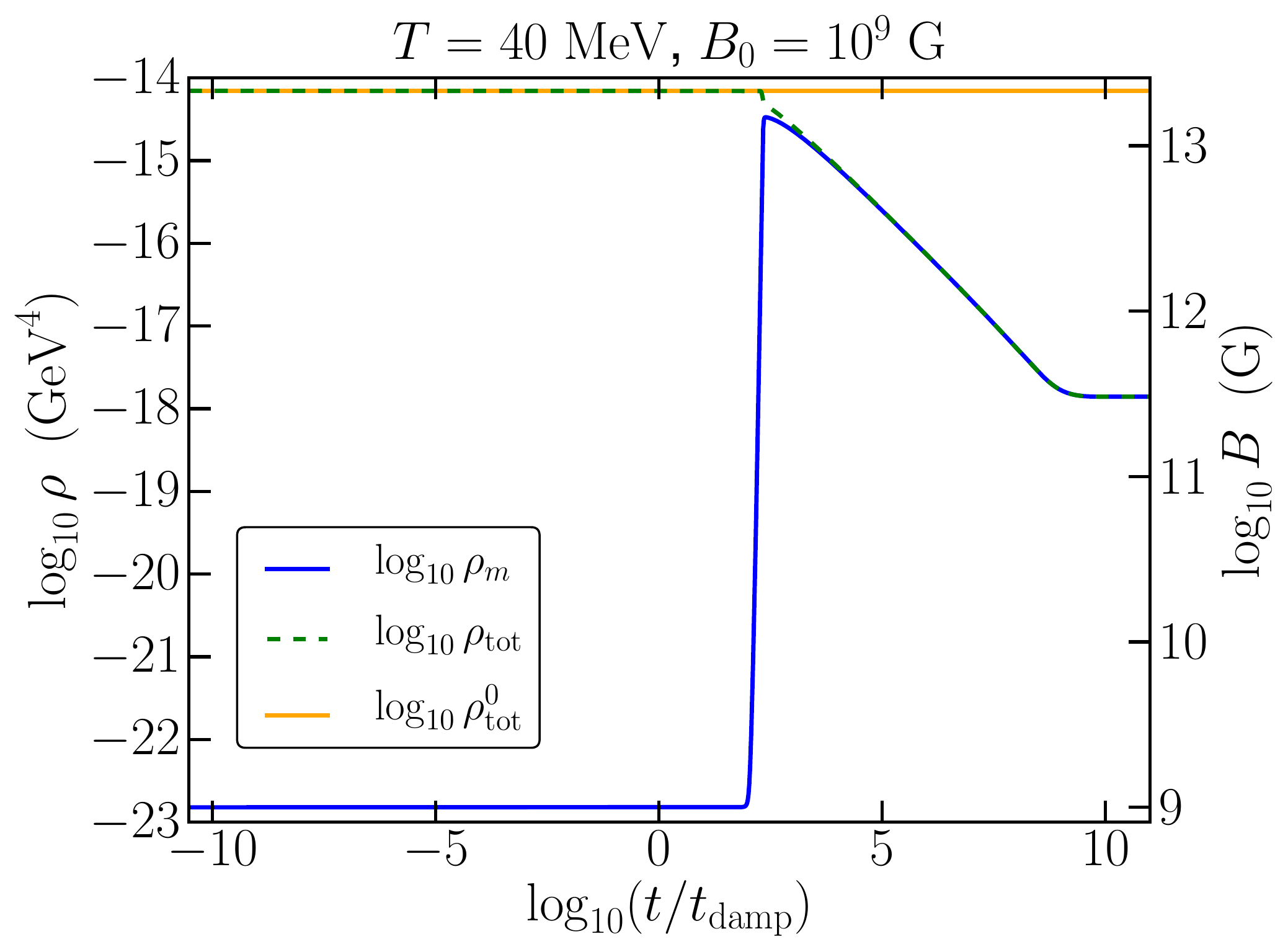}
\hfill
\includegraphics[width=.47\textwidth]{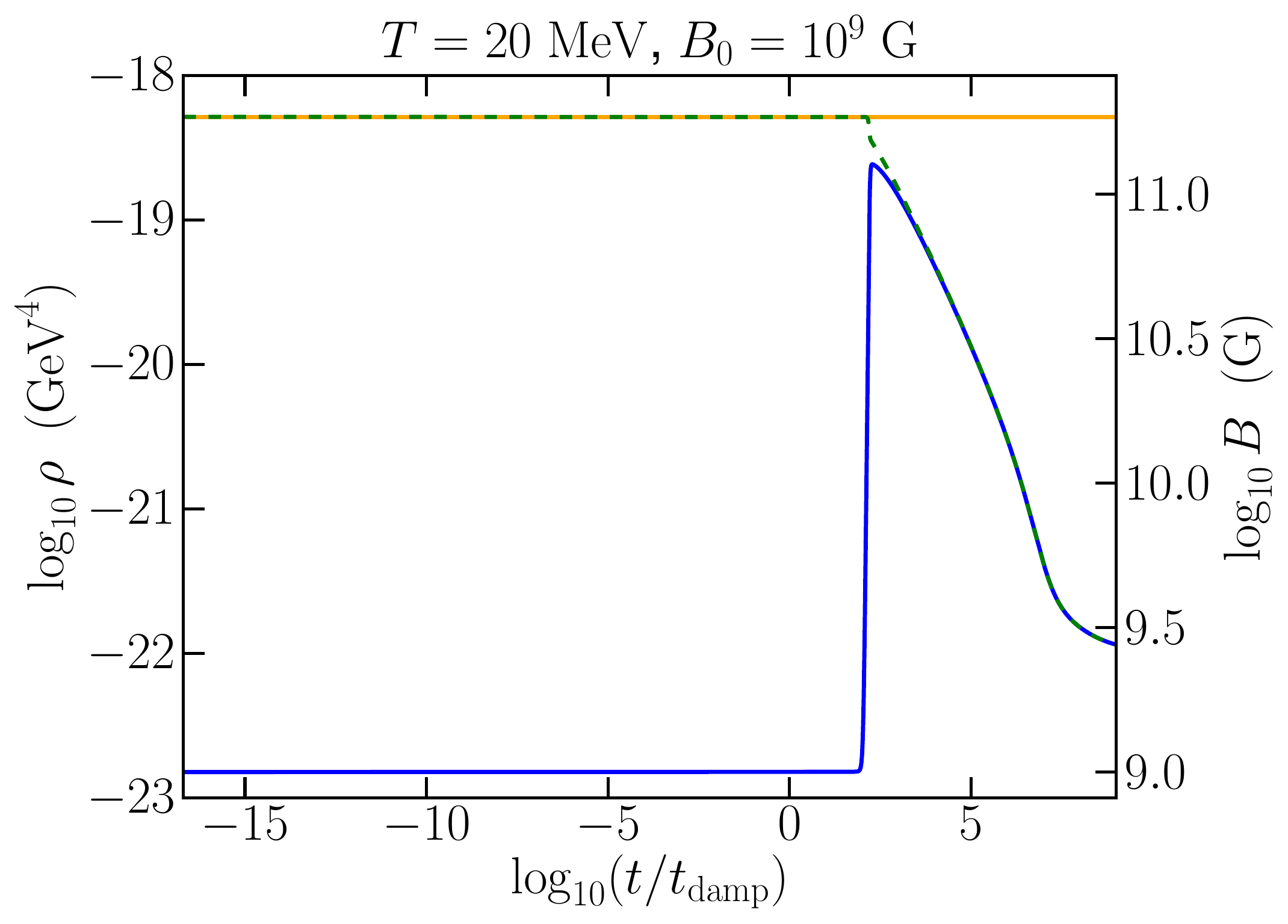}
\caption{\label{pic:over}  Energy densities obtained when the factor $\delta\rho/\rho =0.1$ is included, accounting for the density fluctuations in the neutron star core.}
\end{figure}

Let us finally turn our attention to the case involving density fluctuations $\delta\rho/\rho = 0.1$ in the core of the neutron star, as illustrated in fig.~\ref{pic:over} for seed magnetic fields of $10^{9}$ G. For $T=40$ MeV one obtains a field amplification of now only $\simeq 1\times 10^{13}$ G, whereas for 20 MeV it yields $\simeq 2\times 10^{11}$ G. 

\subsection{Neutrino sphere and cold neutron star} 
If we consider the typical radius of a neutron star to be 10 km, the density at the neutrino sphere is $\sim 10^{11}{\rm g \, cm}^{-3}$. The lepton fraction can be roughly taken as $Y_L\sim 0.1$ \cite{Sulaksono:2013hwa} and the average neutrino energy is at most 16 MeV \cite{Janka}, which translates to a temperature of $T \sim E_\nu/3 \sim 5$ MeV. At this and lower temperatures, an additional particle is required for electron capture to occur, similarly to the modified URCA process $N + p + e_L \rightarrow N + n + \nu_{e_L}$, where $N$ can be either a neutron or a proton. The estimate \eqref{eq:mu5b} gives a chiral asymmetry of $|\mu_{5,b}|\sim 10^{-12}$ MeV which in turn through \eqref{eq:Bmaxdelta} gives negligible maximal magnetic fields.
 
Let us now briefly concentrate on mature neutron stars of about $10^5$ years old, where the core temperature drops to typically $2\times 10^8$ K and the main cooling mechanism is surface photon emission \cite{raffelt}. 
The URCA rate yields an also very small instability equilibrium value $|\mu_{5,b}| \sim 10^{-32}$ MeV, which in turn corresponds to a field growth only around extremely small $k$ and in extremely large time scales, indicating that the chiral asymmetry is also not effective in generating magnetic fields for this regime. We, therefore, obtain significant magnetic field amplification only during the hot initial phase of the neutron star. This is in contrast to \cite{1410.6676,1503.04162} which included the difference in the forward scattering amplitudes for left and right chiral electrons which gives rise to a constant potential term $V_5$. These authors effectively introduced the rescaling $\mu_5 \rightarrow \mu_5 + V_5$ in the terms involving the magnetic field but not in the terms for the chiral asymmetry evolution due to interactions with the background, yielding an artificial steady source of magnetic energy. However, in our opinion this rescaling should be performed in all terms so that it only introduces a constant shift of $\mu_5$ which can be eliminated by redefining $\mu_5$. In other words, the thermodynamic equilibrium abundances can only depend on the total chiral energy difference $\mu_5+V_5$, and not only on $\mu_5$. As a result, $V_5 $should affect neither the scale nor magnitude of the magnetic field amplification. In contrast, if only the kinetic part of $\mu_5$ would be given by \eqref{eq:mu5b}, substituting $\mu_{5,bi}\to\mu_{5,bi}+V_5$ in
\eqref{eq:maxrhom} and using $\mu_{5,bi}\simeq\mu_{5,b}$ with \eqref{eq:mu5b} would yield
\begin{equation}
 \rho_m \lesssim c(T, \mu_e)\frac{\Gamma_f}{\Gamma_w}\left(0.09\frac{\Gamma_w}{\Gamma_f}\frac{T^3}{c(T, \mu_e)}+ V_5\right)^2\,.
\end{equation}
But since $\Gamma_w \propto G_F^2$ and $V_5 \propto G_F$ this would not vanish in the limit of $G_F \rightarrow 0$,
as is expected since if parity is conserved and there is no chiral asymmetry, the energy associated with the chiral asymmetry
should vanish. This indicates that the evolution of the system should not depend on $V_5$. The consideration of a term of this kind has been also discussed and discarded in \cite{1409.3602, Vilenkin} under similar reasoning.  
In Ref.~\cite{Dvornikov:2015ena} the authors claim that the maximal magnetic field energy density
$B^2_{\rm eq}/(8\pi)$
is given by the thermal energy of the nucleons and electrons but no derivation is given and it is unclear
how this is related to $V_5$. Furthermore, saturation of magnetic field growth is introduced ad hoc by substituting
$\mu_5+V_5\to(\mu_5+V_5)/(1+B^2/B^2_{\rm eq})$ without derivation.

\section{Summary and conclusions}
\label{sec:conclusions}
In a supernova core collapse electron capture creates an imbalance between left- and right- handed electrons $\mu_5\sim\,$eV -- keV. It has
been suggested that the chiral anomaly can transform the energy associated with this chiral imbalance into the growth of
helical magnetic fields, possibly up to the high values that have been observationally inferred for neutron stars and magnetars.
In the present work we investigated this possibility within a semi-analytical approach with specific emphasis on the evolution of the total energy which can only decrease or stay constant and thus limits the maximally possible magnetic field strength.
While neutrinos are trapped in the core, density fluctuations allow for local thermodynamic disequilibrium between URCA and inverse URCA rates due to neutrino free-streaming on sufficiently small length scales $\ell_\nu\sim$ cm, which prevents this imbalance to be washed out by the inverse reactions.
For length scales $10^{-6}\,{\rm cm}\lesssim\pi/(2e^2|\mu_5|)\lesssim \ell_\nu\lesssim15\,$cm the chiral magnetic effect can then create
magnetic fields of roughly maximal helicity on time scales short compared to the evolution of the neutron star
before they saturate due to the limited energy associated with the chiral lepton asymmetry.

For a core temperature of 40 MeV, we obtain maximal magnetic field strengths $B_{\rm max}\sim 10^{14}$ G on tens of nanometer length scales
reached within microseconds. 
For lower temperatures, such as 20 MeV, the magnetic fields are smaller and concentrated on larger length scales, and the growth rates are lower.
This suggests that the range of field strengths and power spectra due to the chiral magnetic instability depend on the initial 
temperature of the protoneutron star. The generated fields are not strong enough to account for typical magnetar field strengths
and tend to be produced on submillimeter length scales rather than dipolar fields on the linear scale of the star.
We also find that outside the neutrino sphere, as well as in a cold neutron star at temperatures below $\simeq10$ MeV,
the chiral instability can not lead to significant field amplification,
due to the fact that lower temperatures imply smaller asymmetry values $\mu_5$.

We briefly summarize the differences between our study and other recent work on the chiral magnetic instability in neutron stars. Maximum surface fields of $10^{18}$ G were estimated in Ref.~\cite{1402.4760} by considering a very high and constant $\mu_5=200$ MeV. 
The procedure in Ref.~\cite{1409.3602} yields a chiral asymmetry $\sim 10^{-12}$ MeV (for $T=30$ MeV), several orders of magnitude lower than our result motivated by the electroweak electron capture rate. The responsible mechanism for the magnetic field growth in Ref.\cite{1410.6676,1503.04162,Dvornikov:2015ena} is stated as being due to a potential term $V_5$ that accounts for the parity asymmetric forward scattering of chiral electrons and nucleons and which acts on a much longer time scale, being relevant for cold neutron stars. In our treatment any chiral asymmetry in the forward scattering of electrons on background species does not separately contribute to the enhancement of the magnetic field because only the total asymmetry energy $\mu_5+V_5$ should enter the evolution equations, as also concluded in Ref.~\cite{Vilenkin,1409.3602}.
We rather believe that the magnetic field evolution
only depends on the asymmetry between left- and right-handed electron abundances which in turn is a function of the ratio of
electroweak URCA and spin flip rates.

One important approximation taken throughout this work was neglecting the role that turbulence may play in suppressing the chiral instability by setting ${\bf v}=0$ in the MHD equation. This should be a good
approximation as long as the velocity field is sufficiently smooth on the instability length scales or the power index sufficiently large, as shown in \ref{app:turb}, which
allows to transform into an inertial frame moving along with the plasma. 
We also consider the temperature and the resistivity of a protoneutron star to be constant over the first initial stage of evolution after the supernova collapse. The later should be a good approximation since for $T\gtrsim20$ MeV
all relevant time scales, including the instability growth scale, are short compared to the scale on which the temperature changes, which is a few seconds \cite{burrows86}.
Lower temperatures imply instability time scales longer
than a few seconds on which the cooling of the protoneutron star should be taken into account.
The resistivity that we employed in this work assumes that particles in the protoneutron star core are degenerate whereas a semi-degenerate regime is more realistic. This affects the resistive damping rate and thus the damping time and the timescale in which the instability grows.
Additionally, since the instability timescale is in any case small compared to the dynamical timescale, the final state, in particular the finite magnetic field strength, is not influenced by this uncertainty because it is determined by saturation of the magnetic field energy at a value comparable to the energy in the chiral asymmetry and given by eq.~(\ref{eq:rhomax}).

\acknowledgments
We thank V.~Semikoz for useful discussions.
This work was supported by the ''Helmholtz Alliance for Astroparticle Physics (HAP)'' funded by the Initiative and Networking Fund of the Helmholtz Association and by the Deutsche Forschungsgemeinschaft (DFG) through the Collaborative Research Centre SFB 676 ''Particles, Strings and the Early Universe''.

\appendix
\section{The role of turbulence} \label{app:turb}

In this section we show under which conditions the assumption that the fluid turbulence can be neglected is a good approximation. Comparing the first and third terms of \eqref{eq:Bevol} provides us with an estimate of how large the velocity in the core of a protoneutron star has to be
  to dominate over the chiral term in the MHD equation studied in the present work. 
Let us assume a fluid velocity spectrum of the form
\begin{equation}
\langle \boldsymbol{\upsilon}^2(T, k) \rangle  = \boldsymbol{\upsilon}_i^2(T)\left[\frac{k}{k_i(T)}\right]^n,
\end{equation}
with $k_i$ being the inertial wavenumber. In terms of the length scale $\ell=2\pi/k$ and of the root-mean-square velocity $\boldsymbol{\upsilon}_{\rm rms}= \sqrt{\langle \boldsymbol{\upsilon}^2(T, k) \rangle}$, this gives us for the velocity flow
\begin{equation}
 \boldsymbol{\upsilon}_\ell = \boldsymbol{\upsilon}_{\rm rms}\left(\frac{\ell}{L}\right)^{n/2}, 
\end{equation}
with $L$ the integral length scale and $n$ the power index.
From the MHD equation with the chiral anomaly \eqref{eq:Bevol}, the velocity and anomalous term can be estimated as
\begin{equation}
 \begin{split}
  \boldsymbol{\nabla}\times (\boldsymbol{\upsilon} \times \mathbf{B}) &\sim \frac{\upsilon B}{\ell}, \\
  \frac{e^2}{2\pi^2\sigma}\mu_5  \boldsymbol{\nabla}\times  \mathbf{B} &\sim \frac{e^2 \mu_5 B}{2\pi^2\sigma \ell}, 
\end{split}
\end{equation}
such that the relative importance of the first is roughly dictated by
\begin{equation} \label{eq:turbest}
 \frac{\boldsymbol{\nabla}\times (\boldsymbol{\upsilon} \times \mathbf{B})}{e^2/(2\pi^2\sigma)\mu_5  \boldsymbol{\nabla}\times  \mathbf{B}} \sim 
 2\sigma L \upsilon_{\rm rms} \left[\left(\frac{e}{\pi}\right)^2 L \mu_5\right]^{-(n/2+1)},
\end{equation}
by considering that the relevant scale for the instability is $\ell = 2\pi/k_5= (\pi/e)^2|\mu_5|^{-1}$.

We take the example of a protoneutron star with $T=40$ MeV, with a corresponding conductivity of $\sigma \simeq 0.21$ GeV computed from \eqref{eq:cond}, that according to table \ref{tab:val}  has an electron chiral chemical potential of $2\times 10^{-3}$ MeV. Considering the scale at which most energy will be concentrated after the hot cooling phase being $L\sim$ km, for a Kolmogorov velocity spectrum ($n=2/3$), we obtain
\begin{equation} \label{eq:turbsimp}
 \frac{\boldsymbol{\nabla}\times (\boldsymbol{\upsilon} \times \mathbf{B})}{e^2/(2\pi^2\sigma)\mu_5  \boldsymbol{\nabla}\times  \mathbf{B}} \simeq 5
\times 10^3 \upsilon_{\rm rms}.
\end{equation}
This implies that for $n=2/3$, the effect of the chiral instability dominates unless the fluid velocity of the protoneutron star core is $\upsilon_{\rm rms}\gtrsim 10^{-4}$. As an upper limit to the typical velocities implied we consider the example of $4\times 10^8$ cm/s \cite{Mao:2015nxa}, rendering $\upsilon_{\rm rms} \simeq 10^{-2}$, which indicates that in this regime our results do not apply for Kolmogorov turbulence, but only for larger power indices, such as $n=4/3$ in this case. For a lower limit on the fluid velocity we take the example of $10^5$ cm/s \cite{Link:2011jd}, giving $\upsilon_{\rm rms} \simeq 3\times 10^{-6}$, which according to \eqref{eq:turbsimp} shows that the MHD equation will be dominated by the chiral anomaly for a Kolmogorov type spectrum in this case.

The velocity term in the MHD equation becomes less important as $n$ increases: for a Kraichnan type spectrum ($n=1$), \eqref{eq:turbest} is of the order unity and smaller for $\upsilon_{\rm rms} \lesssim 10^{-3}$. The previous estimates make clear that turbulence can be neglected for relatively small fluid velocities or large power indices, but that depends on the details of the velocity spectrum at play in the core of the protoneutron star.

\end{document}